\documentclass[aps,prb,showpacs,reprint]{revtex4-1}

\usepackage{amsmath}

\newcommand{\kvec}{\boldsymbol{k}}
\newcommand{\nvec}{\boldsymbol{n}}
\newcommand{\mvec}{\boldsymbol{m}}
\newcommand{\qvec}{\boldsymbol{q}}

\newcommand{\deltavec}{\boldsymbol{\delta}}
\newcommand{\cdagger}{c^{\dagger}}

\usepackage{graphicx}

\begin{document}

\title{Tunable graphene bandgaps from superstrate mediated interactions.}

\author{J.P. Hague}
\affiliation{Department of Physical Sciences, The Open University, Walton Hall, Milton Keynes, MK7 6AA, United Kingdom}

\date{9th September 2011}


\begin{abstract}
A theory is presented for the strong enhancement of
graphene-on-substrate bandgaps by attractive interactions mediated
through phonons in a polarizable superstrate. It is demonstrated that
gaps of up to 1eV can be formed for experimentally achievable values
of electron-phonon coupling and phonon frequency. Gap enhancements
range between 1 and 4, indicating possible benefits to graphene
electronics through greater bandgap control for digital applications,
lasers, LEDs and photovoltaics through the relatively simple
application of polarizable materials such as SiO$_2$ and Si$_3$N$_4$.
\end{abstract}




\maketitle

\section{Introduction}

A key goal for graphene research is the development of applications
which require substantial bandgaps, such as digital
transistors. Graphene monolayers have zero band gap, but small gaps
have been observed when graphene is placed on substrates such as SiC
(250meV) \cite{zhou2007a}, gold on ruthenium (200meV)
\cite{enderlein2010a} and predicted for graphene on boron nitride
(100meV) \cite{giovannetti2007a}. A technique for enhancing these gaps
up to the 1eV order of magnitude seen in silicon is crucial to the
development of digital electronics. Moreover, the ability to make
spatially dependent changes to the gap of a semiconducting material
should open a route to new electronic devices and applications, and the
availability of strongly tunable gaps is important to the development
of laser diodes, LEDs, photovoltaics, heterojunctions and photodetectors.

Here, I investigate gap enhancement effects due to interactions
mediated through superstrates placed on graphene systems where a gap
has been opened with a modulated potential. For example, sublattice
symmetry breaking has been suggested for the gap opening mechanism for
the graphene on ruthenium system \cite{enderlein2010a}. The specific
aim is to establish if the small gap size of e.g. graphene on
ruthenium can be increased using attractive electron-phonon coupling
to vibrations in a strongly polarizable superstrate. In low
dimensional materials, strong effective electron-electron interactions
can be induced via interaction between electrons confined to a plane
and phonons in a neighboring layer which is polarizable
\cite{alexandrov2002a}. In the context of carbon systems, experiment
has shown that electron-phonon interactions between carbon nanotubes
and SiO$_2$ substrates have strong effects on transport properties
\cite{steiner2009a} and theory has shown that similar interactions
account for the transport properties of graphene on polarizable
substrates \cite{fratini2008a}.

Besides graphene on substrates, several alternative options have been
put forward to generate gaps in graphene. McCann and Falko
\cite{mccann2006a,mccann2007a} proposed that bilayer graphene develops
a gap when it is gated with an electic field, and the bilayer graphene
gap has been observed experimentally by Ohta {\it et al.} using ARPES
\cite{ohta2006a} and by Zhang {\it et al.} using infrared spectroscopy
\cite{zhang2009a}.

Electron confinement in quasi-1D structures has also been put forward as a
solution to the generation of bandgaps. Graphene nanoribbons with
certain edge orientations were theoretically hypothesised some time
ago to posess gaps \cite{fujita1996a,nakada1996a} and it has been
shown using {\it ab-initio} calculations that reduction of the
nanoribbon width can lead to substantial gap sizes due to electron
confinement, although for gap sizes of the order of an electronvolt,
very narrow nanoribbons with widths of the order of 10 \AA\ are
required. Increase in nanoribbon gaps up to around $300$meV due to
electron confinement has been measured by Han {\it et al.}
\cite{han2007a} although the width variation of the ribbons is
large. Recent developments have allowed for the manufacture of high
quality nanoribbons by unzipping of nanotubes \cite{kosynkin2009a},
and theoretical predictions have been confirmed experimentally, with a
measured gap of $23.8 \pm 3.2$ meV for an (8,1) nanoribbon
\cite{tao2011a}.

A more extreme solution involves changing the chemistry of graphene
monolayers. The generation of large band gaps in graphene on the order
of several eV has been hypothesized for chemical modification with
hydrogen (graphane) \cite{sofo2007a,boukhvalov2008a} and fluorine
(fluorographene) \cite{charlier1993a}. While forms of graphane
\cite{elias2009a} and fluorographene \cite{cheng2010a} have been
manufactured, and hints of bandgaps have been found, there is
currently no consensus on the size of the bandgap, as it is difficult
to obtain even coverage of the hydrogen or fluorine. Also, it is
likely that the wide bandgaps are too large for many digital
applications. Finally, I briefly mention that graphene on some
substrates forms Moir\'{e} patterns, which also lead to a modified
electronic structure \cite{macdonald2011a}.

This paper is organized as follows: a model
for graphene sandwiched between a dielectric superstrate and a
gap-opening substrate is introduced in Sec. \ref{sec:model}. In
Sec. \ref{sec:high} results from Hartree--Fock theory in the high
phonon frequency limit are presented. Perturbation theory and results
for low phonon frequency are presented in Sec. \ref{sec:low}. A
summary and outlook can be found in Sec. \ref{sec:summary}.

\begin{figure}
\includegraphics[width = 85mm]{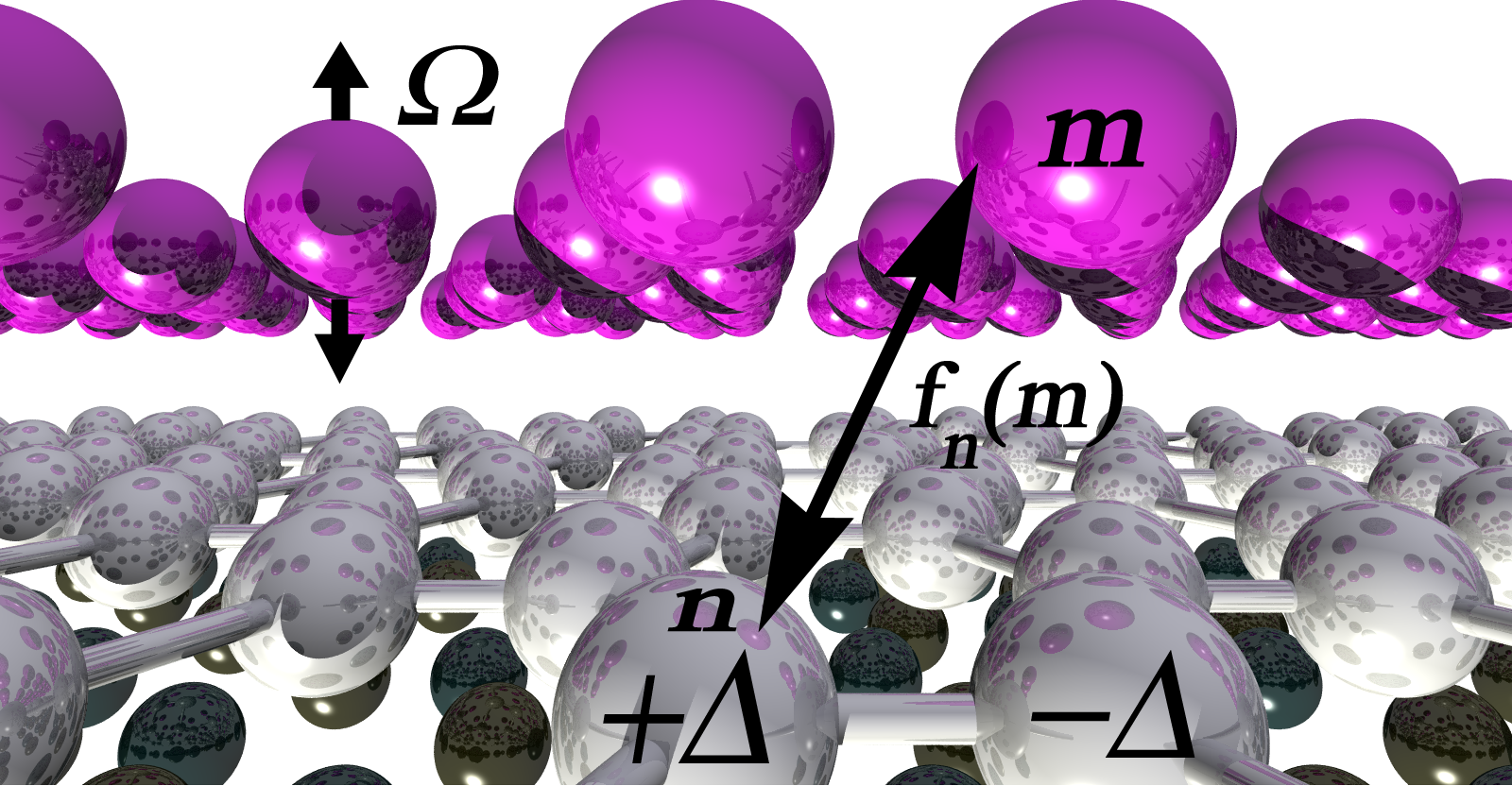}
\caption{(Color online) Graphene-substrate-superstrate system
  annotated with interactions and sublattices. Electron-phonon
  interactions between the graphene layer and superstrate are poorly
  screened, and large interactions of strength $f_{\nvec}(\mvec)$ are
  possible. Ions in the superstrate oscillate with frequency
  $\Omega$. A sites have energy $+\Delta$ and B sites $-\Delta$ due to
  the substrate. Atoms in the graphene sheet are indexed with vector
  $\nvec$ and ions in the superstrate with
  $\mvec$. \label{fig:summary}}
\end{figure}

\section{Model}
\label{sec:model}

A model Hamiltonian for the electronic properties of graphene on a substrate modified by the presence of a superstrate should
have at least three components. The first is a mechanism for electrons
to move through the material. The second is an electron-phonon
interaction and the third a static potential to describe symmetry breaking between graphene sub-lattices by the substrate. The presence of superstrates also opens the possibility of long-range electron-phonon interactions. As such, a
model Hamiltonian for graphene on a substrate has the form,
%
%
\begin{eqnarray}
H & = & -t\sum_{\langle \nvec,\nvec'\rangle\sigma}(a^{\dagger}_{\nvec\sigma}c_{\nvec'\sigma} + c^{\dagger}_{\nvec'\sigma}a_{\nvec\sigma}) - \sum_{\nvec\mvec\sigma} f_{\nvec}(\mvec)n_{\nvec\sigma}\xi_{\mvec} \nonumber\\
&& + \sum_{\mvec} \hbar\Omega (N_{\mvec}+1/2) + \sum_{\nvec\sigma}\Delta_{\nvec}n_{\nvec\sigma}.
\end{eqnarray}
The interactions between electrons in a graphene monolayer and
polarizable ions in the superstrate are shown schematically in
Fig. \ref{fig:summary}. The first term in the Hamiltonian describes
the kinetic energy of tight binding electrons hopping in the graphene
monolayer with amplitude $t$, written in momentum space
as $\sum_{\kvec}(\phi_{\kvec} a^{\dagger}_{\kvec}c_{\kvec} +
\phi^{*}_{\kvec} c^{\dagger}_{\kvec}a_{\kvec})$, where $\phi_{\kvec}=-t\sum_i
\exp(i\kvec.\deltavec_i)$ and $\deltavec_i$ are the nearest neighbor
vectors from graphene A to B sites, $\deltavec_1=a(1,\sqrt{3})/2$,
$\deltavec_2=a(1,-\sqrt{3})/2$, $\deltavec_3=(-a,0)$. Electrons are
created on graphene A sites with the operator $a^{\dagger}_{\nvec}$
and B sites with $c^{\dagger}_{\nvec'}$ and the vectors $\nvec$ are to
atoms in the plane.

The next term in the Hamiltonian describes the electron-phonon
interaction, which has the momentum space form $\sum_{\kvec\qvec}
g_{\kvec \qvec}
[c^{\dagger}_{\kvec-\qvec}c_{\kvec}(d^{\dagger}_{\qvec} + d_{-\qvec})
  + a^{\dagger}_{\kvec-\qvec}a_{\kvec}(b^{\dagger}_{\qvec} +
  b_{-\qvec})] +\sum_{\kvec\qvec} \tilde{g}_{\kvec \qvec}
[a^{\dagger}_{\kvec-\qvec}a_{\kvec}(d^{\dagger}_{\qvec} + d_{-\qvec})
  + c^{\dagger}_{\kvec-\qvec}c_{\kvec}(b^{\dagger}_{\qvec} +
  b_{-\qvec})]$. Phonons are created in the superstrate above A sites
with $b^{\dagger}_{\mvec}$ and above the B sublattice with
$d^{\dagger}_{\mvec}$, so the displacement $\xi_{\mvec}\propto
(b^{\dagger}_{\mvec}+b_{\mvec})$ for site A. The interaction strength
$f_{\mvec}(\nvec)=\kappa/[(\mvec-\nvec)^2+1]^{3/2}$ has the classic
Fr\"ohlich form ($\kappa$ is a coupling constant). The effective
phonon-mediated interaction between electrons can be characterised
by the function $\Phi(\nvec,\nvec') =
\sum_{\mvec}f_{\mvec}[\nvec]f_{\mvec}[\nvec']$. A complication of this
Hamiltonian is that a basis of two atoms is needed to represent the
honeycomb lattice, which is the reason for two interactions; $g$ and
$\tilde{g}$. For local electron-phonon coupling, $\tilde{g}$ vanishes
and $g$ becomes momentum independent. For simplicity, I study this
simplified version of the electron-phonon Hamiltonian
\cite{covaci2008a}, i.e the effective interaction is approximated as a
Holstein model which has local interaction,
$\Phi(\nvec,\nvec')=\delta_{\nvec,\nvec'}$. The third term represents
the energy of phonons with frequency $\Omega$, where $N_{\mvec}$ is
the number operator for phonons. It is appropriate to mention the
effect of the electron-phonon interaction on suspended graphene, which
simply leads to renormalizations of phonon and electron modes
\cite{castroneto2009a,covaci2008a}.

The Fr\"ohlich form for the electron-phonon interaction has been
demonstrated experimentally for carbon nanotubes on SiO$_2$
\cite{steiner2009a}, and it has been established to be the main
scattering mechanism for graphene on SiO$_2$ \cite{fratini2008a}. At
weak electron-phonon coupling, the effects of Holstein and Fr\"ohlich
interactions are qualitatively similar on two-dimensional lattices
\cite{hague2006a} and the Holstein form is used throughout this
paper. It is worth noting that there may be quantitative changes to
the results presented in this paper for longer range interactions.

For completeness, I briefly discuss an alternative class of
electron-phonon interaction where phonon motion couples to the
electron hopping, leading to a Su--Schrieffer--Heeger (SSH) style
interaction which can lead to dimerization (and could act against the
mechanism used here) \cite{su1980a}. To have a strong SSH interaction,
a material must be very flexible (e.g. the SSH model is very good for
describing polymers). Here, since the graphene is sandwiched between two
other materials, the material will be held rigid and in-plane SSH
interactions are expected to be much smaller than the Holstein style
interaction considered here.

I also note that there may be some modulation of the strength of the
electron-phonon interaction due to incommensurability of the
superstrate. The effects of this incommensurability are
straightforward to estimate in 1D, by computing the sum
$\Phi(\nvec,\nvec') = \sum_{\mvec}f_{\mvec}[\nvec]f_{\mvec}[\nvec']$
for values of $\nvec$ intermediate to the lattice points. It is found that the interaction strength
varies by around $\pm 8\%$ of the average value if the superstrate is
incommensurate. It is not expected that modulations of this magnitude
will qualitatively change the results presented here.

To complete the model of the graphene-substrate-superstrate system,
the final term in the Hamiltonian describes interaction between
electrons and a static potential, $\Delta_{\nvec}$, induced by the
substrate. In particular a modulated potential where A sites have
energy $\Delta$ and B sites $-\Delta$ leads to breaking of the
symmetry between A and B sites and gives rise to a gap. In the
following, I examine the effects of electron-phonon interaction
on this gap. The effect of phonons on substrate induced gaps has not
previously been studied, and as I show, the implication for the
gap is significant.

\begin{figure}
\includegraphics[width = 85mm]{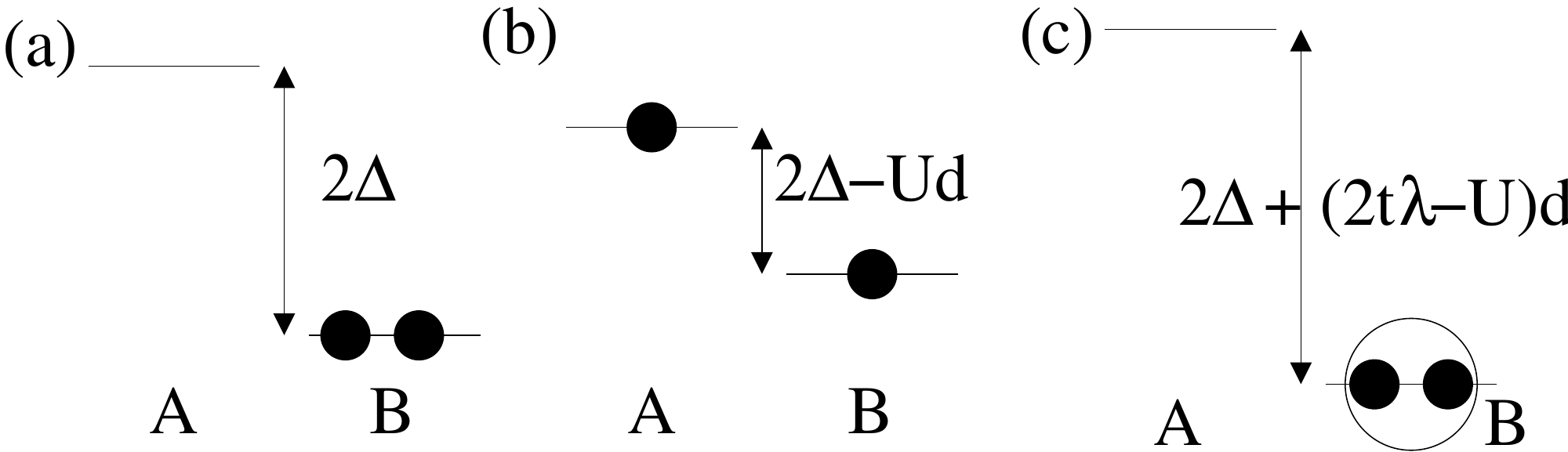}
\caption{Schematic of the physical processes highlighted in the large phonon frequency limit. (a) The modulated potential with magnitude $\Delta$ opens a gap. (b) The effect of Coulomb repulsion is to stop both electrons sitting on the same site, effectively closing the gap. (c) The attractive phonon mediated electronic interaction with coupling constant $\lambda$ pulls electrons onto the same site and reduces the hopping, effectively enhancing the gap. $d$ is the difference in electron occupation between sites A and B.}
\label{fig:schematic}
\end{figure}


\section{High phonon frequency}
\label{sec:high}

To illustrate the core physical content of the model, I initially study the large phonon frequency limit where a Lang-Firsov canonical transformation can be used to derive an effective Hubbard Hamiltonian \cite{lang1963a},
\begin{equation}
H = \sum_{\kvec} \epsilon'_{\kvec} \cdagger_{\kvec B \sigma}c_{\kvec A \sigma} + {\rm H.c.} + \sum_{i}U'n_{iA\uparrow}n_{iA\downarrow} + \sum_{i}\Delta_{i} n_{i}
\end{equation}
Where $U' = U-2t\lambda$, $|\epsilon'_{\kvec}|^2 = 3t'^2 + t'^2(2\cos(k_y\sqrt{3})+4\cos(k_y\sqrt{3}/2)\cos(3k_x/2))$ is the dispersion for the graphene lattice, $t' = t \exp(-t\lambda/\Omega)$ and $\lambda=\Phi(0,0)/2tM\Omega^2$ is the dimensionless electron-phonon coupling which is expected to be smaller than unity ($M$ is the ion mass). A local Coulomb repulsion $U$ has been included for completeness, although the effects of this in the graphene monolayer are limited to renormalization of the electron bands since no phase transition (to e.g. a Mott insulator) is measured, avoiding the need for more complicated treatments of the Coulomb repulsion. Fig. \ref{fig:schematic} shows the basic physical processes that lead to gap modification in graphene on substrate systems in the large phonon frequency limit. Panel (a) shows the gap induced from the modulated potential, where electrons prefer to sit on the lower energy site B. If sufficient Coulomb repulsion is applied, electrons sit on different sites, and this leads to an effective lowering of the gap as shown in panel (b). Contrary to this, an electron-phonon interaction leads to lower energies when electrons are on the same site and also tends to decrease the effective hopping, which is expected to increase the gap (panel c).

\begin{figure}
\includegraphics[width = 85mm]{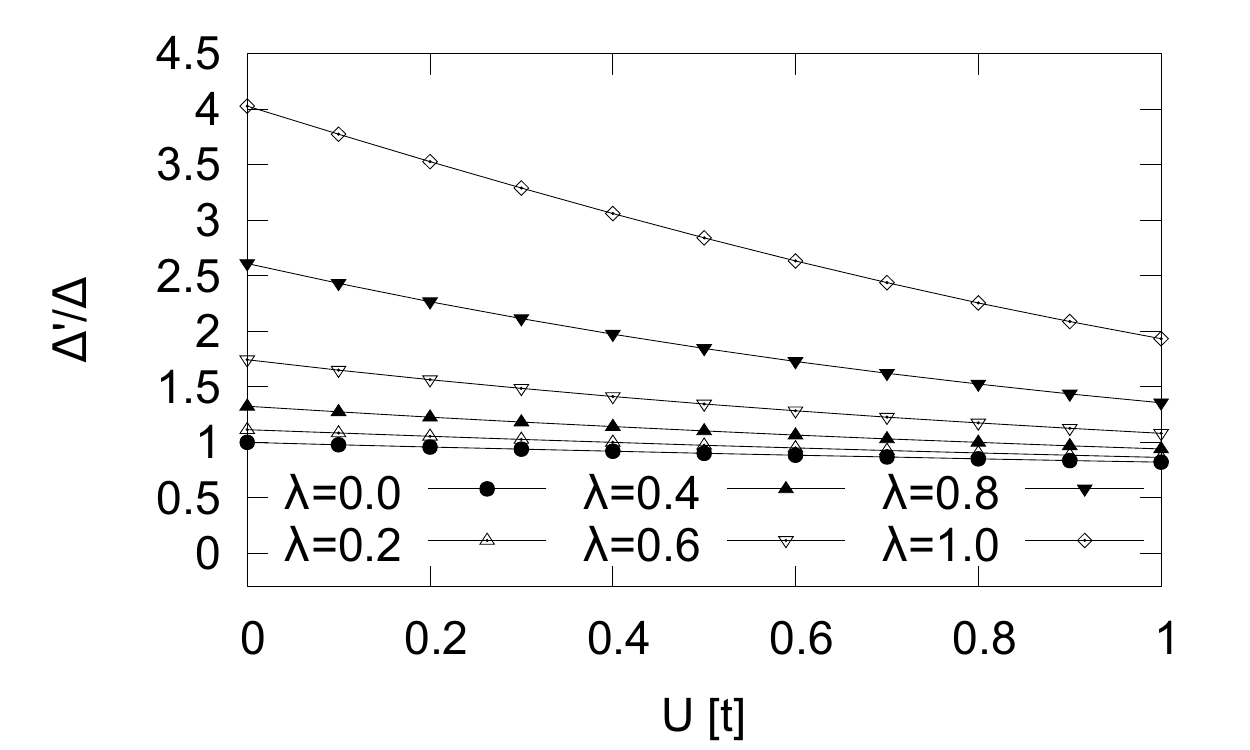}
\caption{Enhancement of the substrate induced graphene bandgap in the anti-adiabatic (large phonon frequency) limit. Enhancement factors of $\sim 4$ can be seen for medium sized $\lambda$. Here, parameters $T=0$, $\hbar\Omega=t$, $\Delta=0.1t$ are used. The results here are illustrative of the core physics, and solutions of full Eliashberg style equations for physically realistic parameters are shown in Fig. \ref{fig:eliashberg}.}
\label{fig:hartreefock}
\end{figure}

For weak coupling, the Hamiltonian may be decoupled using the standard Hartree-Fock scheme, i.e. $H\approx U'\langle n_{iA\uparrow} \rangle n_{iA\downarrow} + U'\langle n_{iA\downarrow} \rangle n_{iA\uparrow} - U'\langle n_{iA\uparrow} \rangle \langle n_{iA\downarrow} \rangle$. A mean field solution is then taken, with $\langle n_{iA} \rangle = n+d$ and $\langle n_{iA} \rangle = n-d$. In the following, the system is half-filled. Minimizing the total energy with respect to $d$, a gap equation for $d$ is obtained,
\begin{equation}
d = - \frac{1}{V_{BZ}} \int {\rm d}^2\kvec \frac{(Ud/2-t\lambda d+\Delta)/2}{\sqrt{|\epsilon_{\kvec}'|^2+\Delta'^2}},
\end{equation}
where $\Delta' = (U-2t\lambda)d/2 + \Delta$ is the effective bandwidth
once interactions are taken into account, and $V_{BZ}$ is the
Brillouin zone volume. This may be solved by using a binary
search. The results for $\Delta'$ can be seen in
Fig. \ref{fig:hartreefock}. The effect of increased Coulomb repulsion
is a small decrease in the effective gap. The effect of increased
electron-phonon coupling $\lambda$ is far more dramatic, and gap
increases are seen for all $\lambda$ at any $U$ value. I note that the
enhancement increases as the bare gap, $\Delta$, decreases. Crucially,
enhancement factors of $\sim 4$ can be seen for medium sized $\lambda$
which could increase the moderate gaps seen experimentally in graphene
on substrate systems to the $\sim 1$eV gap sizes of silicon and
germanium, with the caveat that changes in the form of the
electron-phonon could quantitatively change this result. By changing
the form of the superstrate or substrate, significant control could be
exercised over the graphene band gap.

\section{Low phonon frequency}
\label{sec:low}

Given the large gap enhancement in the large phonon frequency limit, it is appropriate to examine if the enhancement is present at low phonon frequency. It is not obvious that the intuitive gap enhancement seen at large phonon frequencies should still occur at low frequencies where retardation effects could lead to large pairs that are not localized. For low phonon frequency and weak coupling, low order perturbation theory can be applied. I derive a set of self-consistent equations assuming the following ansatz for the self energy,
\begin{equation*}
\mathbf{\Sigma}(i\omega_n) \approx \left(
\begin{array}{cc}
i\omega_n(1-Z_n)+\bar{\Delta}_n & 0 \\
0 & i\omega_n(1-Z_n)-\bar{\Delta}_n
\end{array}
\right).
\end{equation*}
Here, the momentum independent form of the ansatz (local approximation) is reasonable because of localization by the modulated potential $\Delta$ and the electron-phonon interaction. Off diagonal terms are absent because they do not feature in the lowest order perturbation theory. The quasi-particle weight, $Z_n$, is shorthand for $Z(i\omega_n)$ and $\bar{\Delta}_n$ is the gap function. Matsubara energies for Bosonic quantities are $\omega_s = 2\pi k_B T s$ and for Fermions are $\omega_n = 2\pi k_B T (n + 1/2)$, where $T$ is the temperature and $n$ and $s$ are integers.

The non-interacting graphene Green function in the presence of a modulated potential has the form,
\begin{equation}
\mathbf{G}_0^{-1}(\kvec,i\omega_n) = \left(
\begin{array}{cc}
i\omega_n - \Delta & \phi_{\kvec}^{*} \\
\phi_{\kvec} & i\omega_n + \Delta
\end{array}
\right).
\end{equation}
The full Green function can be established using Dyson's equation $\mathbf{G}^{-1}(\kvec,i\omega_n) = \mathbf{G}_0^{-1}(\kvec,i\omega_n) - \mathbf{\Sigma}(i\omega_n)$, leading to,
\begin{equation}
\mathbf{G}^{-1}(\kvec,i\omega_n) = \left(
\begin{array}{cc}
i\omega_n Z_n - \Delta - \bar{\Delta}_n & \phi_{\kvec}^{*} \\
\phi_{\kvec} & i\omega_n Z_n + \Delta+\bar{\Delta}_n
\end{array}
\right).
\end{equation}

Substituting the expression for the Green function into the lowest order contribution to the self energy,
\begin{equation}
\Sigma_{ii}(\kvec,i\omega_n) = -Tt\lambda\sum_{i\omega_{s}}\int \frac{{\rm d}^2\qvec}{V_{BZ}} G_{ii}(\kvec-\qvec,i\omega_{n-s})d_0(\qvec,\omega_s).
\end{equation}
Here, the phonon propagator, $d_0(\qvec,\omega_s)=\Omega^2/(\Omega^2+\omega_s^2)$. The off diagonal elements of the lowest order self energy are zero in the case of local interaction, so they do not feature in the ansatz. Since there is a modulated potential, it is necessary also to take into account the tadpole diagram (which has a contribution since the effects of the modulated potential can not simply be absorbed into the chemical potential) leading to the following Eliashberg style equations for $\bar{\Delta}_n$ and $Z_n$,
\begin{equation}
\bar{\Delta}_n = 2t\lambda\delta n - t\lambda k_B T \sum_s \int {\rm d}\epsilon \frac{D(\epsilon)\, \Delta'_{n-s}d_0(i\omega_s)}{\omega_{n-s}^2Z^2_{n-s}+\Delta'^2_{n-s}+\epsilon^2},
\end{equation}
\begin{equation}
\delta n = k_B T \sum_n \int {\rm d}\epsilon \frac{D(\epsilon)\, \Delta'_{n}}{\omega_{n}^2Z^2_{n}+\Delta'^2_{n}+\epsilon^2}
\end{equation}
and
\begin{equation}
Z_n = 1-\frac{t\lambda k_B T}{\omega_n} \sum_s \int {\rm d}\epsilon \frac{D(\epsilon)\,\omega_{n-s}Z_{n-s}d_0(i\omega_s)}{\omega_{n-s}^2Z^2_{n-s}+\Delta'^2_{n-s}+\epsilon^2},
\end{equation}
where $\delta n$ is the difference between the density of electrons on sites A and B and the full gap is $\Delta'_n = \bar{\Delta}_n+\Delta$.  Here, the density of states for graphene in the absence of a gap, $D(\epsilon)$, has the form given in Ref. \onlinecite{castroneto2009a}. Note that these gap equations differ from those for a superconductor.

The equations may be solved self consistently by performing a truncated sum on Matsubara frequencies. The maximum Matsubara frequency was kept constant with the value, $\omega_{\rm max}=75t$ which is sufficiently large to ensure that asymptotic behavior of the gap function was achieved. The longest self-consistent solutions took around 2 weeks, increasing as $1/T$ and $\lambda$.
%
%
The following parameters have been chosen to match graphene with a modest bare gap of the order of magnitude seen in the systems discussed in the introduction: $\Delta=0.05t$ corresponding to a gap of around 280meV, realistically achievable phonon energies of $\hbar\Omega=0.01t=28$meV, $\lambda<1$ and temperatures on the order of room temperature $k_{B}T=0.01t$ corresponding to $\sim 324$K. The dynamical quasi-particle weight, $Z_n$ is of order unity for all parameters considered here and the gap equation has very weak frequency dependence at the values of $\Omega$ that have been considered.

\begin{figure}
\includegraphics[width = 85mm]{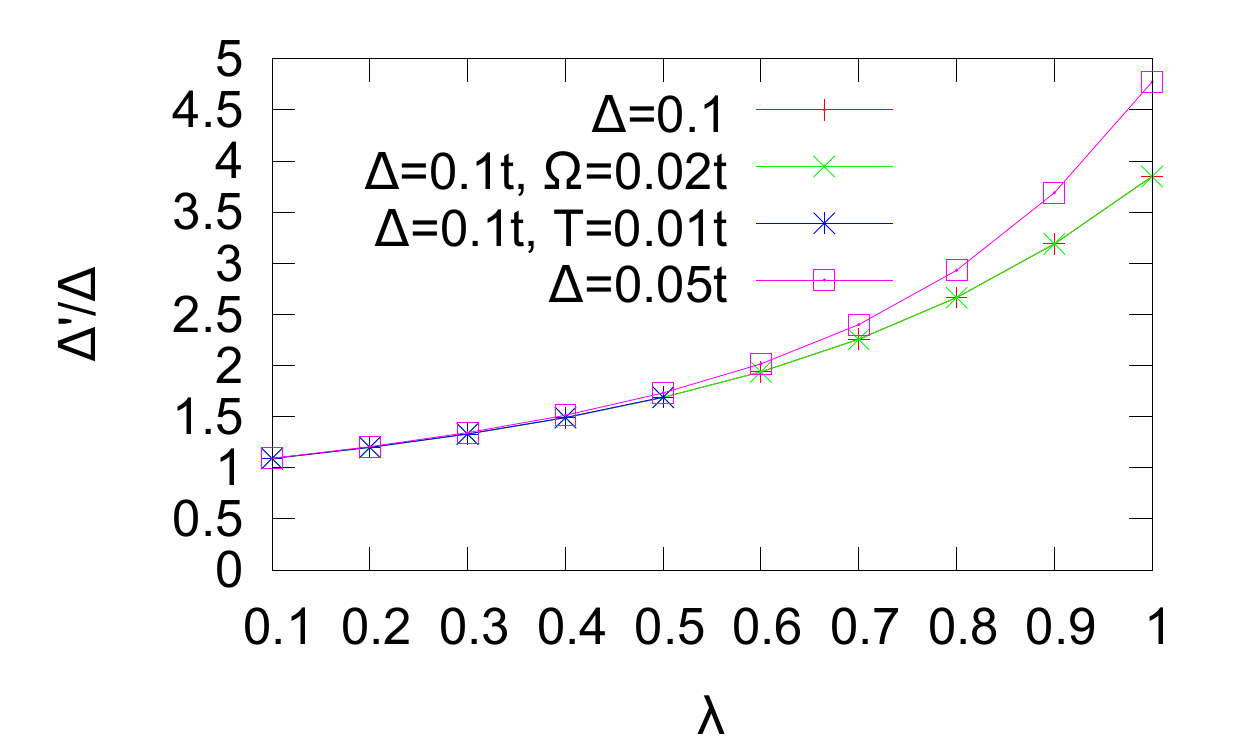}
\caption{(Color online) Enhancement of the substrate induced graphene bandgap in the adiabatic (low phonon frequency) limit. Gap enhancements of up to 4 are readily achievable, and for smaller $\Delta$ the gap enhancement is more pronounced at larger $\lambda$. Realistic parameters are used: $\Delta=0.1t$ and $\Delta=0.05t$ corresponding to bare band gaps of $2\Delta = 0.56$eV and $0.28$eV respectively, $t=2.8$eV, $\hbar\Omega=0.01t=28$meV and $0.02t=56$meV and $k_B T = 0.02t=56$meV and $0.01t=28$meV corresponding to $T=648$K and $324$K and $\lambda\le 1$. In the plot $k_BT=0.02t$ and $\hbar\Omega=0.01t$ unless otherwise specified. All energy scales are much less than the band-width and change in $T$ and $\Omega$ has only a very small $<$1\% effect on the gap. For such a low phonon frequency, the gap function $\Delta_n$ has a very weak frequency dependence. The frequency dependence gets even smaller as temperature drops.}
\label{fig:eliashberg}
\end{figure}

The gap enhancement factor $\Delta'/\Delta$ is shown in
Fig. \ref{fig:eliashberg}. Significant gap enhancement can be seen,
with a swift rise at intermediate $\lambda$ and achieving a factor 3
at around $\lambda = 0.8$. The enhancement factor increases slightly
with decreasing $\Delta$ but is essentially unchanged by modifications
to phonon frequency and temperature for the parameter values used
here. I note that there may be some quantitative reduction to this
enhancement for the longer range Fr\"ohlich interaction. Consideration
of electron-phonon couplings of up to the order of unity are
reasonable since even in metals such as lead, the electron phonon
coupling can be large ($\lambda\sim 1.55$) \cite{reinert2003a} and in
the absence of interplane screening, even larger couplings should be
possible. Therefore, gap sizes relevant to digital graphene devices
should be achievable by placing ionic superstrates on top of e.g. the
graphene on rubidium system, and such a system could make a good
starting point for experimental investigations of the gap
enhancement. It is noted that since rubidium is a conductor, an
insulating material which leads to the same undressed gap would be
necessary to make a working digital device. It is possible to rule out
a number of materials which have been used for top and bottom gating
in transport measurements. I note that gaps can also be induced in
bilayer graphene \cite{mccann2006a} and graphene nanoribbons
\cite{brey2006a}, so a similar mechanism of substrate mediated
electron-phonon interaction may enhance gaps in those systems.

\section{Summary and conclusions}
\label{sec:summary}

In summary, I have presented a theory for the enhancement of graphene
band-gaps by polarizable superstrates. The theory predicts gap
enhancements of up to 4 times from electron-electron interactions
mediated through phonons in a polarizable ionic superstrate. The
generation of sizable graphene bandgaps is a key problem for the use
of graphene in many technologically important applications. The theory
shows that the relatively simple addition of polarizable superstrates
to graphene systems can provide a way of making large enhancements to
graphene gaps. I suggest the following recipe for experimental
investigation of an enhanced graphene gap: Form SiO$_2$ or Si$_3$N$_4$
layers on top of graphene on ruthenium intercalated with gold. I note,
however, that for digital applications an insulating substrate is
required, so if the effect can be verified experimentally it will be
necessary to find additional substrates that cause gaps in the
graphene spectrum. It is hoped that this work will stimulate
experiment, leading to tunable gapped graphene systems with
applications in digital electronics, LEDs, lasers and photovoltaics.

\section{Acknowledgments}

I acknowledge EPSRC grant EP/H015655/1 for funding and useful
discussions with P.E. Kornilovitch, A.S. Alexandrov, M. Roy,
P. Maksym, E. McCann, V. Fal'ko, N.J. Mason, N.S. Braithwaite,
A. Ilie, S. Crampin and A. Davenport.

\bibliography{graphene_sub_analytic}

\end{document}